\documentclass[aps,notitlepage,twocolumn,nofootinbib,floatfix,pra,10pt]{revtex4-2}
\usepackage{amsmath,amsfonts,amssymb,amsthm,bbm,graphicx,enumerate,times,mathtools,physics,marvosym,wasysym}
\usepackage{hyperref}
\hypersetup{colorlinks,linkcolor=blue,citecolor=red,urlcolor=blue}
\usepackage{csquotes}
\usepackage{float}
\usepackage[ruled,longend]{algorithm2e}
\usepackage{fancyref}
\usepackage{subcaption} 
\usepackage{afterpage} 
\usepackage{placeins} 
\usepackage{caption} 
\usepackage{etoolbox}
\usepackage{mdframed}
\usepackage{subcaption}
\usepackage{tabularx}
\usepackage{appendix,soul}
\usepackage[dvipsnames]{xcolor}
\usepackage[noabbrev,capitalise]{cleveref}
\usepackage{enumitem}
\begin{document}

\title{ArtA: Automating Design Space Exploration of Spin Qubit Architectures}




\author{Nikiforos Paraskevopoulos$^{1,2}$}
\author{David Hamel$^{1}$}
\author{Aritra Sarkar$^{1,2}$}
\author{C. G. Almudever$^{3}$}
\author{Sebastian Feld$^{1,2}$}

\affiliation{$^{1}$Quantum and Computer Engineering Department, Delft University of Technology, 2628 CD Delft, The Netherlands}
\affiliation{$^{2}$QuTech, Delft University of Technology,  2628 CJ Delft, The Netherlands}
\affiliation{$^{3}$Computer Engineering Department, Universitat Politècnica de València, Camino de Vera, s/n, 46022 València, Spain}

\begin{abstract}


In the fast-paced field of quantum computing, identifying the architectural characteristics that will enable quantum processors to achieve high performance across a diverse range of quantum algorithms continues to pose a significant challenge. Given the extensive and costly nature of experimentally testing different designs, this paper introduces the first Design Space Exploration (DSE) for quantum-dot spin-qubit architectures. Utilizing the upgraded \textit{SpinQ} compilation framework, this study explores a substantial design space comprising 29,312 spin-qubit-based architectures and applies an innovative optimization tool, ArtA (\textit{\textbf{Art}}ificial \textit{\textbf{A}}rchitect), to speed up the design space traversal. ArtA can leverage 17 optimization configurations, significantly reducing exploration times by up to 99.1\% compared to a traditional brute-force approach while maintaining the same result quality. After a comprehensive evaluation of best-matching optimization configurations per quantum circuit, ArtA suggests specific as well as universal architectural features that provide optimal performance across the examined circuits. Our work demonstrates that combining DSE methodologies with optimization algorithms can be effectively used to generate meaningful design insights for quantum processor development.
\end{abstract}

\maketitle

\section{Introduction}

As the field of quantum computing rapidly advances, different qubit technologies exhibit unique hardware and performance characteristics. It still remains to be seen which one (e.g., superconducting, trapped ions, quantum dots, photonics, defect-based on nitrogen-vacancy diamond centers) will succeed in scaling up quantum computing systems with high-quality qubits \cite{resch2019quantum,chatterjee2021semiconductor}. Among them, spin qubits in quantum dots emerge as a compelling avenue for achieving scalability for practical quantum computation \cite{chatterjee2021semiconductor}. To this day, spin qubits are at an early stage in their development, with Intel's Tunnel Falls chip currently boasting the highest count of twelve spin qubits \cite{intel2023quantum}. Despite this, their inherent scalability advantages suggest that a robust two-dimensional design \cite{vandersypen2017interfacing,li2018crossbar, hendrickx2021four,borsoi2024shared,zhang2025universal,john2024two,unseld2024baseband} could be scaled up relatively easily once fabrication techniques and quality improve up to a certain level. Designing a chip, however, involves multiple architectural design choices whose impact in the future can only be fully assessed through experimental studies post-scaling the technology. Exploring all these design possibilities simultaneously can be expensive, time-intensive, and impractical. Therefore, simulating how different design choices can affect performance in a range of quantum applications is crucial to assist the development. This process facilitates highlighting architectural insights that can guide the technology forward and will allow quantum researchers to make more informed decisions, streamline efforts, and hasten the development of these devices.

Recognizing the intricate challenges of designing quantum processor architectures, this study initiates the first Design Space Exploration (DSE) for quantum dot spin-qubit architectures, both from a current technological perspective and a future one. Therefore, in this work, we have identified a wide range of representative architectural features and abstracted them into usable input variables, resulting in $29,312$ different architectures. To facilitate this exploration, we have enhanced the compilation capabilities of \textit{SpinQ} compilation framework \cite{SpinQ} to handle all these input variables and updated the definition of the Estimated Success Probability (ESP) metric to include not only operational errors but also crosstalk and decoherence errors. These transformations, and many more, establish \textit{SpinQ} as the first compilation and DSE framework for spin qubit architectures.


Standard DSE processes \cite{pimentel2022methodologies,mathew2019rramspec}, however, will be impractical time-wise, especially for such large spaces, as they rely on a brute-force approach to traverse and subsequently analyze the design space. To address this, our approach employs ArtA (\textit{\textbf{Art}}ificial \textit{\textbf{A}}rchitect), a built-in tool containing multiple optimization methods for automating the DSE process of \textit{SpinQ}, thus taking less time than the brute-force approach with the same quality of results. The abilities of ArtA are (a) to suggest which of the seventeen optimization configurations can find the architecture with the desired ESP the fastest compared to the brute-force approach and (b) which architectural design characteristics are key for building high-performance spin-qubit devices per quantum circuit. 

Our results demonstrate ArtA's ability to compare all optimization techniques for each quantum circuit and obtain a solution up to 99.1\% faster, on average, than brute-forcing. Then, equipped with these insights, we move on to conduct a DSE analysis of $29,312$ spin qubit architectures and provide valuable insights into best design practices. Firstly, we highlight the critical need to maximize the parallelization of quantum gates rather than minimizing crosstalk between qubits. Secondly, we find that the communication method via shuttle operations is achieving higher ESP in large-scale circuits than SWAPs, but for single-qubit gates, pulse-based rotations are preferred over shuttle-based ones. Finally, a universal recommendation based on all tested circuits indicates that the likelihood of success increases when prioritizing the parallelization of single-qubit gates over two-qubit gates. This demonstrates that the combination of DSE techniques and optimization algorithms can effectively guide quantum processor designers with meaningful recommendations in the current and future stages of spin-qubit technology. Notably, we also observe that certain quantum circuits achieve optimal performance even with reduced hardware capabilities. This highlights the capability of our approach to uncover optimal architectural configurations without relying on highly complex or resource-intensive hardware designs.

The main contributions of this paper are:
\begin{enumerate}
    \item The design space definition which comprises five main characteristics of current and, potentially, future spin qubit architectures. 
    
    \item The upgraded \textit{SpinQ}, the first compiler and DSE framework for spin qubit architectures. In this version, we have updated the ESP formula to include operational, crosstalk, and decoherence-induced errors.
    
    \item ArtA, the first tool consisting of seventeen optimization method configurations automating the DSE process of spin qubit architectures.
    
    \item Evaluation of best-matching optimization configurations per quantum circuit.

    \item Evaluation of best-matching architectural characteristics per quantum circuit and universally (i.e., for all used circuits) in terms of ESP.
    
\end{enumerate}

The remainder of this paper is structured as follows: In Section \ref{Spin qubits as a scalable platform}, we discuss the scalability potential of quantum-dot spin qubits as well as the importance of addressing engineering challenges while taking into account the performance of the resulting architecture. Then, in Section \ref{Problem Statement}, we motivate the need for an automated DSE analysis for spin qubit devices and formulate four research questions for this work. After that, in Section \ref{SpinQ and the design space}, we define the design space consisting of various input architectural variables and establish the new functionalities of \textit{SpinQ}. Another upgrade in \textit{SpinQ} is the enhanced ESP metric, described in Section \ref{Figure of Merit}, which is used as the figure of merit for our exploration. In Section \ref{ArtA (Artificial Architect)}, we introduce ArtA and two new metrics to evaluate its performance. In the results Section \ref{Results and Evaluation}, we first evaluate ArtA's performance across all used quantum circuits and determine the best optimization method configuration. Then, we conduct a detailed analysis of the best architectural designs for each quantum circuit and form valuable insights. We conclude our work and provide ideas for future directions in Section \ref{Conclusion}. 

\section{Challenges for scaling up spin qubit devices} \label{Spin qubits as a scalable platform}

Spin-qubit technologies are distinguished by their unique physical features, which position them as a highly scalable solution for quantum computing. The advantages of spin-qubits include a significantly smaller size — up to a thousand times less than other qubit technologies — combined with decades of semiconductor manufacturing expertise, long coherence times coupled with short gate durations, and high operational temperatures \cite{yoneda2018quantum,camenzind2022hole,hendrickx2021four,chatterjee2021semiconductor,RevModPhys.85.961,PhysRevA.57.120,vandersypen2017interfacing,veldhorst2015two,zajac2015reconfigurable,watson2018programmable}. At the core of this technology lies the quantum dot, which can contain a trapped electron(s) or hole(s) to form a physical qubit \cite{hanson2007spins}. Spin qubits are manipulated electromagnetically using multiple precision-engineered gate electrodes that facilitate either single- or two-qubit operations through exact timing of pulse sequences across various quantum dot configurations. Studies have expanded these systems into one-dimensional and two-dimensional arrays \cite{hendrickx2021four,borsoi2024shared,zhang2025universal,john2024two,unseld2024baseband}, exploring different structures and material combinations.

Despite the mentioned advantages, spin qubit quantum processors are not as advanced as other qubit technologies in terms of qubit counts and device availability. Major technological hurdles are related to the so-called interconnect bottleneck \cite{vandersypen2017interfacing}, and various fabrication challenges towards scaling up \cite{burkard2023semiconductor,bluhm2019semiconductor,de2023silicon,de2021materials}. On the upside, there have been significant efforts \cite{li2018crossbar,boter2022spiderweb,vandersypen2017interfacing,hill2015surface,franke2019rent,paquelet2020multiplexed,pauka2019cryogenic,veldhorst2017silicon,ivlev2025operating} to tackle these challenges. 

However, solving them can not guarantee successful quantum algorithm executions. This is because the quality and quantity of qubits are not the only factors determining a high-performing quantum processor. The architectural constraints qubits need to comply with in \textit{how} they are operated are equally important. For instance, the benefits of qubits with excellent operational fidelity can easily be outweighed by low qubit connectivity and limited natively supported quantum gates. Similarly, high-quality qubits with low crosstalk interference are not enough when they can not be addressed in parallel. These, and many more, are interlinked architectural trade-offs that affect the actual performance of the quantum processors.

During the Noisy Intermediate-Scale Quantum (NISQ) era \cite{preskill2018quantum}, predicting which architectural features and trade-offs will facilitate successful quantum circuit executions while maintaining reasonable hardware requirements remains challenging. To date, there has not been a systematic study exploring spin-qubit architectures through this prism of providing concrete guidelines for future development. Therefore, in this study, we exploit such an opportunity and alleviate the need for time-consuming, expensive, and technology-dependent experimental studies.

\section{Problem statement} \label{Problem Statement}


The need to apply DSE techniques for designing and optimizing full-stack quantum computing systems, or components thereof, has been emphasized in previous work \cite{almudever2021structured,tomesh2021quantum,beverland2022assessing}. These techniques and similar ones have been successfully applied across different levels of the quantum computing stack \textcolor{black}{which predominantly rely on brute-force exploration (i.e., exhaustive search). This is largely due to the inherently small design spaces in current quantum computing systems, which renders brute-force evaluation computationally feasible. For instance, DSE methodologies} have been used to explore ion-trap quantum processors \cite{murali2020architecting}, evaluate compilation techniques \cite{quetschlich2023compiler} and qubit connectivities \cite{lin2022domain,liang2023superconducting,9229178} for superconducting qubits. Such techniques can abstractly convert architectural characteristics into design variables and, in this way, quantize the design space, expressing current and future design possibilities that otherwise would be impossible or too time-consuming to physically realize. Starting from there \cite{pimentel2022methodologies,mathew2019rramspec}, the steps for a proper DSE: i) Describe the problem in terms of input variables or parameters (design choices), ii) select the performance metrics, iii) choose a global cost function as a figure of merit that combines different performance metrics, and iv) find models (behavioral, analytical, from experimental data) that connect input parameters with metrics or directly to the figure of merit. Applying these DSE techniques allows researchers to uncover performance trends, model current and future multidimensional design spaces, and identify optimal design points across a wide range of application use cases.

Although this methodology is well established and understood in many disciplines, it can be challenging to implement in such early stages for spin-qubit technologies. One of the most difficult aspects is the development of a simulation framework incorporating a range of representative architectural variables with a large-scale perspective. Firstly, it is difficult to predict which architectural features will be relevant for many generations of devices to come, and secondly, the framework itself has to be flexible to incorporate new ones easily.

\textcolor{black}{In practice, performing exhaustive exploration can be highly time-consuming; thus, even in classical systems, the design space is typically constrained to a manageable size by applying simplifying technological assumptions. While optimization algorithms have been successfully employed to efficiently explore significantly larger design spaces in classical computing domains \cite{theocharis2024multi, chen2024distributed, madsen2006multi, palesi2002multi, zhang2008extended, reagen2017case, schafer2009adaptive, wang2006design}, such techniques remain largely unexplored in quantum computing DSE applications. As spin-qubit technology rapidly advances, the design space continues to grow with new considerations, making earlier explorations quickly outdated. Even assuming a brute-force approach already exists, the exponential growth in possible configurations can soon make it impractical to evaluate thousands of designs. Therefore, there is a clear need for an automated framework capable of efficiently managing this DSE process, significantly reducing the exploration time across current and future spin-qubit architectures.}

To capture the challenges of the above problem statement, we ask the following four research questions:

\begin{enumerate} 

    \item What characteristics of spin-qubit architectures are representative, and how can they be incorporated in a DSE framework? Answered in Section \ref{SpinQ and the design space}


    \item How can a DSE framework for investigating preferred spin-qubit architectural characteristics be constructed? Answered in Section \ref{SpinQ and the design space}
    

    \item Which optimization methods are most effective for automating the DSE process for each quantum circuit individually, as well as across all used circuits collectively? Answered in Section \ref{Comparison of optimization methods}

    
    \item Which architectural characteristics in the selected design space are key for building high-performance spin-qubit devices for the specifically given quantum circuits? Answered in Section \ref{Architecture Analysis}
    
    
\end{enumerate}

\section{\textit{SpinQ} and the design space} \label{SpinQ and the design space}

\begin{table}
\centering
    \caption{Summary of values of the considered architecture variables}\label{tab:DesignSpace}
    \includegraphics[width=1\linewidth]{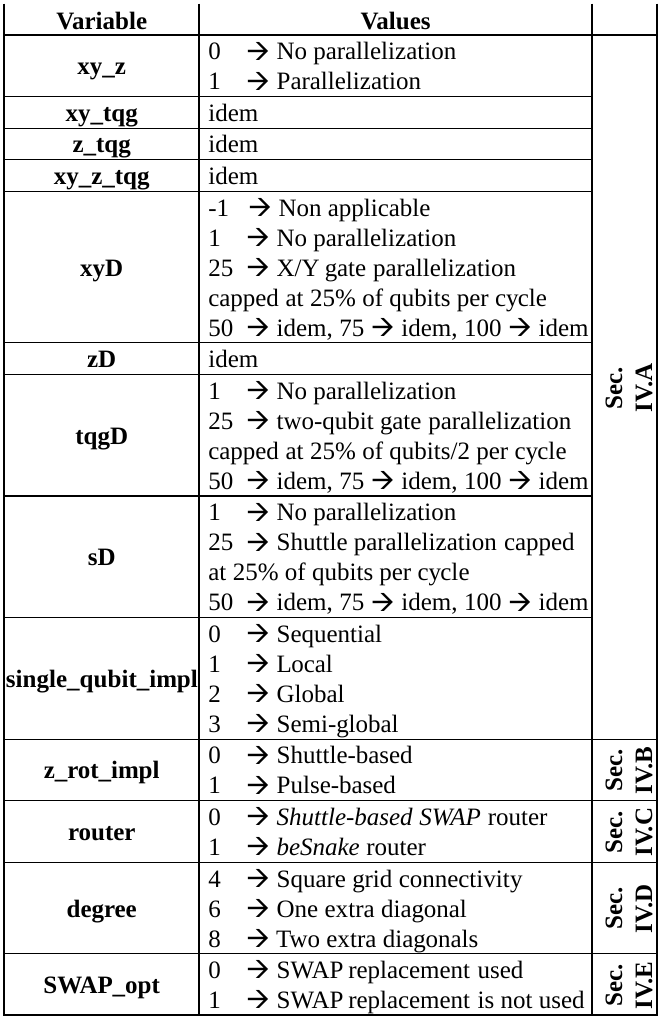}
\end{table}

To answer the first two research questions, we conducted an analysis of relevant spin-qubit architectural properties with a long-term view and distilled them into numerous variables. These properties were selected to be representative of spin qubit devices as well as to be able to form performance-related trade-offs between them. The latter will prevent ``maximizing'' one without causing a performance penalty through another (refer to Section \ref{Figure of Merit}). The defined design space is summarized in Table \ref{tab:DesignSpace}, including their possible values. An architecture is then characterized by one (valid) combination of variable values. Below, we describe the considered design space analytically. It should be noted that the DSE analysis in this work uses either discrete or categorical variables \cite{nardi2019practical}, as we aim to provide performance trends and rankings between optimization methods and architectures.

\subsection{Operational gate constraints} 

The set of gate constraints can be vast; therefore, we decided to abstract them into the following variables that affect which (a) gate types can be executed simultaneously and (b) how many of them can be parallelized at the same time step.

\begin{itemize}
    \item[a)] \textbf{Combinations of gate types that can be parallelized:}
    
    (\texttt{xy\_z, xy\_tqg}, \texttt{z\_tqg}, \texttt{xy\_z\_tqg} = \textit{0, 1})
    
    As the names suggest, these variables are boolean and signify whether specific gate types can be executed in parallel. For example, \texttt{xy\_z} defines if X or Y gates can be executed simultaneously with Z gates.
    
    \item[b)] \textbf{Constraints affecting the parallelization per gate type expressed in percentage}:
    
    (\texttt{xyD}, \texttt{zD}, \texttt{tqgD}, \texttt{sD}  = \textit{-1, 1, 25\%, 50\%, 75\%, 100\% })
    
    These variables indicate how many gates of each type can be parallelized, expressed as a percentage relation to the total number of qubits. For example, \texttt{xyD} = 25 in a 40-qubit architecture means that at most ten X or Y gates can exist in a single cycle. Note that the notion of a cycle refers to the basic unit of time representing one step in a sequence of gates of a quantum circuit, and each step may contain multiple gates. The \texttt{zD} and \texttt{sD} variables indicate the same for Z rotations and shuttle operations, respectively. The \texttt{tqgD} variable indicates the parallelization amount for two-qubit gates. As these always involve two qubits, they are expressed as a percentage of half of the total number of qubits. A value of $-1$ means that the parallelization degree is not user-defined but dictated by a specific single-qubit gate implementation explained next.
    
    \item[c)] \textbf{Single-qubit gate implementation:}
    
    (\texttt{single\_qubit\_impl} = \textit{0 $\rightarrow$ Sequential, 1 $\rightarrow$ Local, 2 $\rightarrow$ Global, 3 $\rightarrow$ Semi-global})
    
    
    This variable indicates in what manner single-qubit gates are carried out. A \textit{Global} single-qubit gate implementation means the same rotation axis and angle are applied to all qubits. For qubits that do not participate in a cycle, locally applied disable instructions prevent their rotation. Then, the \textit{Semi-Global} implementation is considered with the same rotation scheme as in \cite{helsen2018quantum} and \cite{SpinQ}, which is essentially equivalent to the implementations of the crossbar architecture proposed in \cite{li2018crossbar}. In this implementation, qubit rotations are implemented semi-globally, meaning that either all qubits in odd or even column parities can be rotated at a time with the same axis and angle, and unwanted qubit rotations have to be reversed afterward by additional instructions \cite{SpinQ}. A Local implementation enables arbitrary parallel execution of single-qubit gates, allowing any combination of rotation axes and angles to be applied simultaneously within a single cycle. Finally, a \textit{Sequential} single-qubit gate implementation does not allow parallelizations, meaning that every cycle will be occupied by one gate only. It should be noted that depending on the \texttt{single\_qubit\_impl} value, only certain combinations of categories (a) and (b) are possible. For instance, in the \textit{Global} and \textit{Semi-Global} implementations, the \texttt{xyD} variable is not applicable as these implementations predetermine the upper limit of single-qubit gates allowed during a cycle. 
\end{itemize}

\subsection{Implementation of Z rotations:}

(\texttt{z\_rot\_impl} = \textit{0, 1})
    
This boolean variable stipulates the hardware-level implementation of Z rotation gates. The first option is a high-fidelity shuttle-based Z rotation, which is achieved with two time-sensitive qubit shuttles to and from a neighboring column \cite{SpinQ,helsen2018quantum,morais2019mapping,li2018crossbar}. This single-qubit manipulation belongs to the so-called ``hopping spins'' type \cite{wang2024operating}, and it is possible to achieve an arbitrary rotation axis, but in this work, we only assume the aforementioned variety for simplicity. The second option is a regular pulse-based Z rotation, where Z gates are implemented similarly to X/Y gates by magnetic pulse interactions. With the first choice being a unique characteristic of spin-qubits, we are interested in exploring which one of the two provides better performance across many quantum circuits. Similarly to the variables mentioned earlier, only certain combinations are possible depending on \texttt{single\_qubit\_impl} values. For example, with a \textit{Semi-Global} single-qubit gate implementation and a pulse-based Z rotation gate, the \texttt{zD} variable is not applicable for the same reason the \texttt{xyD} variable is not applicable, as we explained before in category (b).

\subsection{Connectivity of the device:}

(\texttt{degree} = \textit{4, 6, 8})

\begin{figure*}
         \centering
     \begin{subfigure}[b]{0.5\textwidth}
         \centering
         \includegraphics[width=\textwidth]{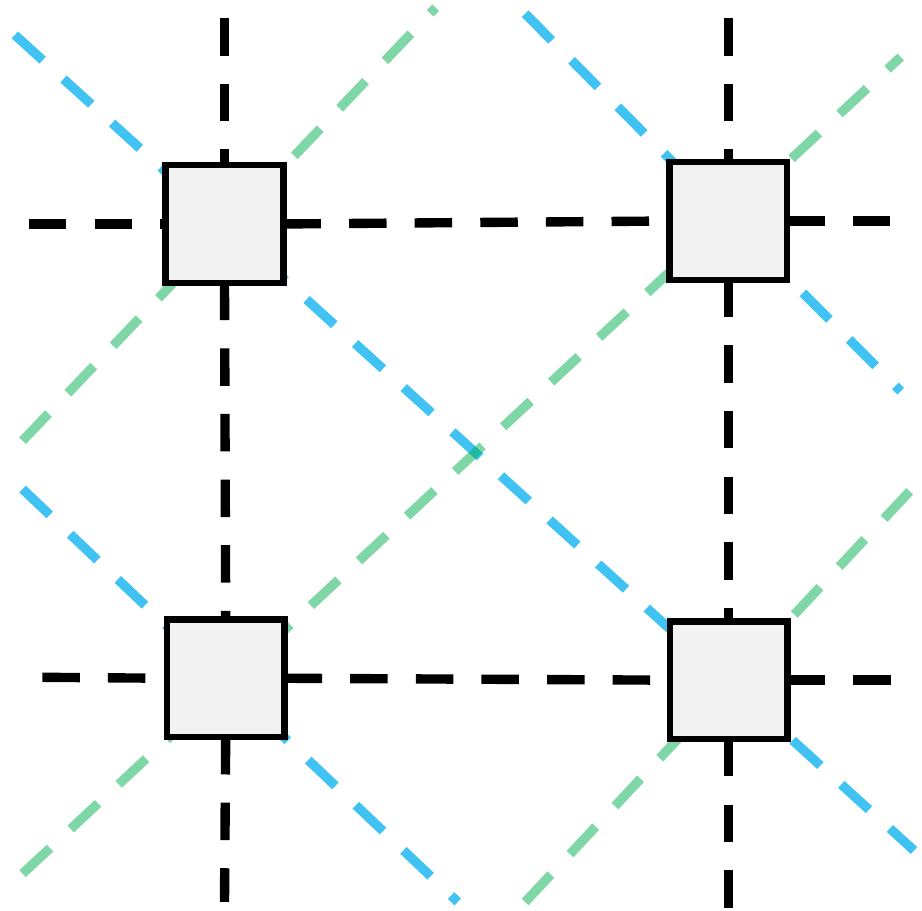}
         \caption{ }
         \label{fig:topologiesa}
     \end{subfigure}
\hfill
     \begin{subfigure}[b]{0.5\textwidth}
         \centering
         \includegraphics[width=\textwidth]{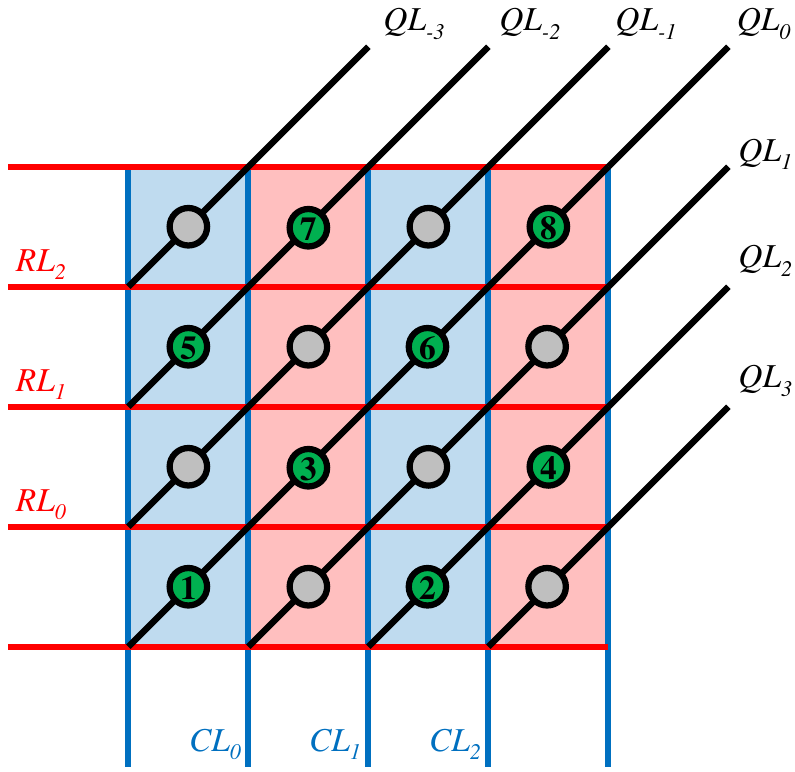}
         \caption{ }
         \label{fig:topologiesb}
     \end{subfigure}
\hfill
     \begin{subfigure}[b]{0.4\textwidth}
         \centering
         \includegraphics[width=\textwidth]{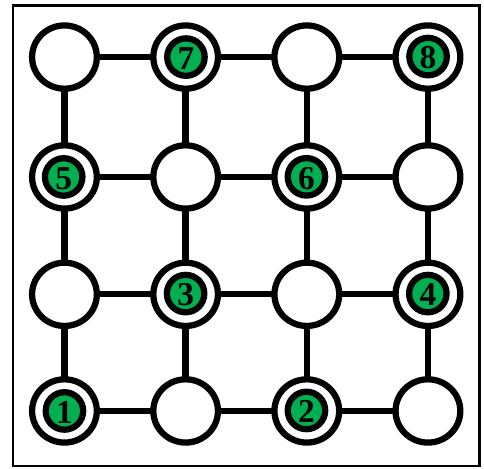}
         \caption{ }
         \label{fig:topologiesc}
     \end{subfigure}
         \caption{ \textcolor{black}{\textbf{(a)} The different coupling graphs considered in the design space. The coupling graph consists of the black line edges when \texttt{degree} = \textit{4},  black and green edges when \texttt{degree} = \textit{6}, and all edges when \texttt{degree} = \textit{8} \textbf{(b)} Schematic overview of the crossbar architecture, taken from \cite{SpinQ}, with the various shared operational control lines: vertical (column line, CL), horizontal (row line, RL), and diagonal (qubit line, QL). These lines are used with precise pulse sequences and shared among multiple sites, sixteen quantum dots in this figure, to perform operations on qubits \cite{li2018crossbar,helsen2018quantum,morais2019mapping} under specific operational constants. Here, the qubits (green circles with numbers) are initialized in a checkerboard pattern. \textbf{(c)} Abstraction of the crossbar architecture taken from \cite{paraskevopoulos2024besnake} representing the coupling graph between qubits with \texttt{degree} = 4. Each circle represents a quantum dot, and each edge represents a coupling link signifying allowed interactions.}}
         \label{fig:topologies}
\end{figure*}

This variable stipulates the considered coupling graphs by addressing them through their average node degree. A coupling graph is a graph of possible direct interactions (edges) between qubits (nodes). The first coupling graph we used is a square grid (\texttt{degree} = 4). The second is identical to the first but adds one diagonal connection in each square (i.e., each qubit is coupled with six neighboring qubits and with three at the corners, \texttt{degree} = 6). Finally, the third adds the second diagonal in each square (i.e., each qubit is coupled with eight neighboring qubits and with three at the corners, \texttt{degree} = 8). \textcolor{black}{Figure \ref{fig:topologiesa} presents the coupling graphs considered with color-coded edges for each of the three types. The actual dimensions of each coupling graph, and thus the overall size of the quantum architecture, are determined by the number of qubits required by the quantum algorithm under execution.}

\textcolor{black}{To further ease the understanding of the architectural landscape of the design space, in Figure~\ref{fig:topologiesb}, we depict the crossbar architecture~\cite{li2018crossbar} as a real-world example of an architecture represented in the design space. In this figure, the operational lines confine sixteen quantum dots, eight of which are occupied by spin qubits. The operational gate constraints in this architecture arise from the shared-control lines and other physical properties as described in \cite{li2018crossbar,helsen2018quantum,morais2019mapping,SpinQ}. Among these constraints, the coupling connectivity is particularly relevant, as it corresponds to a square grid with \texttt{degree} = 4 in this implementation. Consequently, the topology of the crossbar architecture can be abstracted as a two-dimensional grid, as depicted in Figure~\ref{fig:topologiesc}.} 

We have now reviewed all variables that pertain solely to the hardware. Next, we examine two additional variables that influence the compilation process. Specifically, these variables can alter the routing methodology based on the underlying hardware communication constraints.

\subsection{Routing methods via shuttling:}

(\texttt{router} =  \textit{0 $\rightarrow$ Shuttle-based SWAP}, \textit{1 $\rightarrow$ beSnake})

There are two available algorithmic options. The first one, introduced in \cite{SpinQ,morais2019mapping}, is a \textit{shuttle-based SWAP} routing algorithm. In particular, this algorithm was tailored around the unique constraints of the crossbar architecture \cite{li2018crossbar}, which necessitated the maintenance of the checkerboard physical qubit pattern to achieve a time-efficient compilation process. One of the advantages of this algorithm is, in fact, the maintenance of the checkerboard pattern, which in turn keeps the crosstalk interference low and exclusively enables the \textit{Semi-Global} rotations scheme. The second routing option is called \textit{beSnake} \cite{paraskevopoulos2024besnake}, and it is capable of freely shuttling qubits around any topology and in any direction, handling complex routing scenarios involving parallelized gates.

\subsection{SWAP replacement:}

(\texttt{SWAP\_opt} = \textit{0, 1})
   
This binary variable concerns an optional functionality of the \textit{beSnake} routing algorithm, which can replace a sequence of shuttles with a SWAP gate under certain conditions \cite{paraskevopoulos2024besnake}. These conditions are satisfied when the accumulated fidelity of a parallelized shuttle sequence exceeds the fidelity of only executing a SWAP on that particular location. This variable is useful for creating insights for architectures that support SWAPs compared to others that do not support them. 

\subsection{Design interdependencies and verification} \label{Design interdependencies and verification}

\begin{figure*} 
    \centering
    \includegraphics[width=\linewidth]{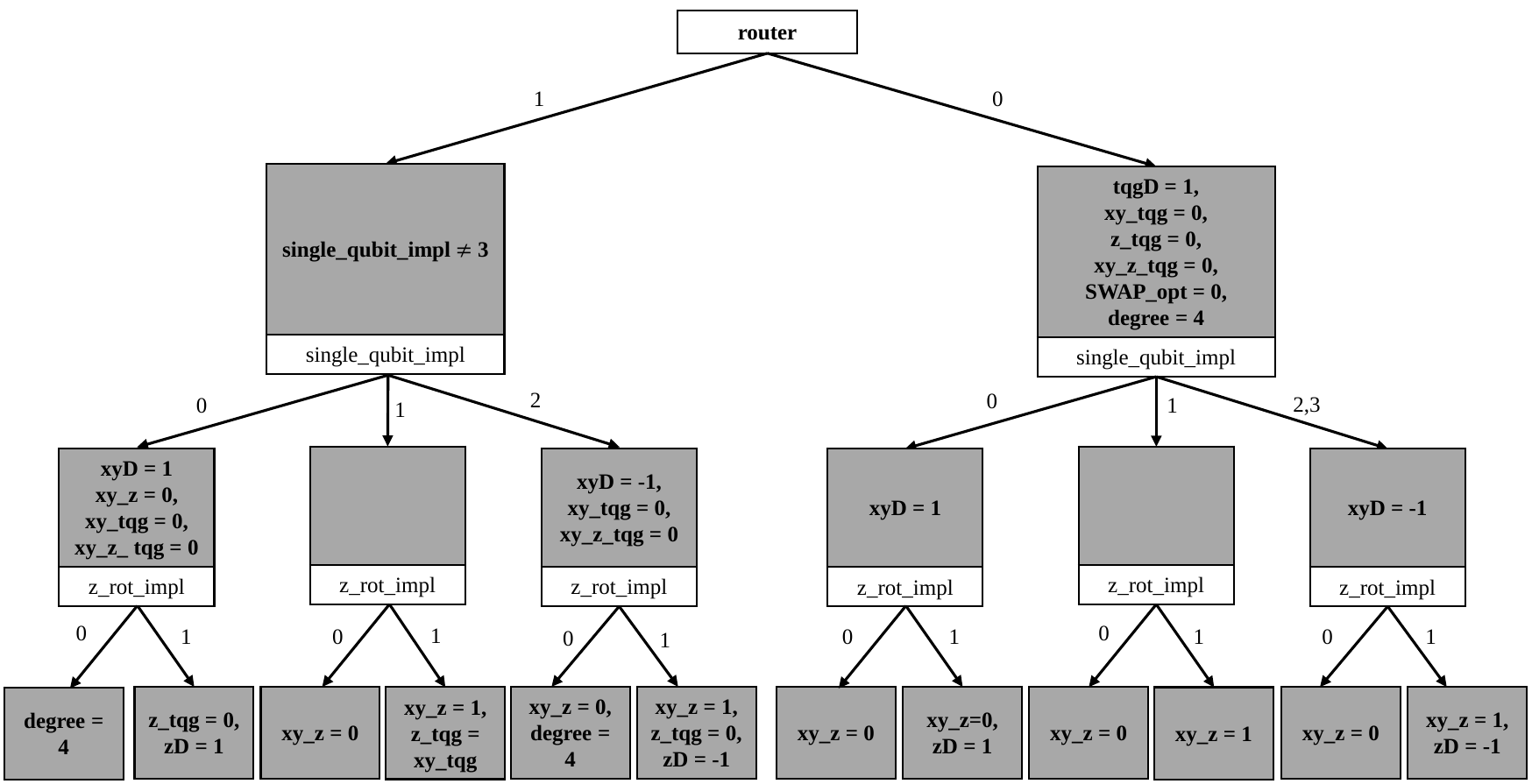}
    \caption{Tree of valid architectural configurations that \textit{SpinQ} handles.}
    \label{fig:Interdependences}
\end{figure*}

All these variable values act as input into the upgraded \textit{SpinQ} housing its new \textit{configurable compiler}, making it the first compilation and DSE framework for spin-qubit architectures. After this upgrade, we are able to provide \textit{SpinQ} with multiple combinations of variable values, compile each input circuit iteratively for all valid architectures, and calculate and store multiple performance metrics, such as the ones used in \cite{SpinQ}. These are gate overhead, circuit depth overhead, and ESP of the compiled quantum circuit(s).

As mentioned, not all combinations of variable assignments are valid, as there are interdependencies between them; hence, \textit{SpinQ} automatically filters the design space to allow only valid combinations. If all combinations were allowed, presented in Table \ref{tab:DesignSpace}, there would be a total of $1,105,920$ architectures, but the valid ones used in this work amount to $29,312$. Figure \ref{fig:Interdependences} shows the interdependences tree that \textit{SpinQ} uses to filter valid architectures. 

Another significant component of \textit{SpinQ} is the compiler verification tool \cite{SpinQ}. This tool is crucial, given the absence of real devices for testing. For this work, the verification functions have been enhanced to handle all valid architectures via our two-step method \cite{SpinQ}, which thoroughly scrutinizes circuits at each compilation stage.

\section{Figure of merit} \label{Figure of Merit}

The selection of an appropriate figure of merit is a critical step to ensure fair and objective evaluation of quantum processor architectures. It must reflect how specific design features influence circuit performance, allowing each architecture to be characterized based solely on its inherent trade-offs and capabilities. To achieve this, each design variable must contribute directly to the final metric, such that the interplay between performance gains and associated costs (e.g., crosstalk or decoherence) is properly captured. This prevents the optimization tool ArtA (introduced in Section \ref{ArtA (Artificial Architect)}) from disproportionately favoring certain architectures by, for instance, maximizing the degree of gate parallelization without accounting for the negative effects it may introduce, such as increased crosstalk. To highlight an example of a design trade-off, we note that the higher the degree of parallelization, the shorter the circuit depth will be, and therefore, the shorter the algorithm execution time, but it will result in higher crosstalk. Such considerations also expresses more accurately design decisions in real experimental quantum processors.


When a quantum processor is not physically available, a common method for estimating its performance is to compute the Estimated Success Probability (ESP) \cite{quetschlich2022predicting,nishio2020extracting,near-term}, derived from the compiled quantum circuit. This method is significantly more computationally efficient than alternatives such as Schrödinger-Feynman simulation, tensor network contraction, or noise model simulations, all of which scale poorly with increasing circuit size and qubit connectivity \cite{SpinQ, near-term}. In this work, we build upon the ESP definition introduced in \cite{SpinQ}, shown in Equation \ref{eq:ESPold}, and extend it to incorporate all previously discussed architectural considerations. The revised figure of merit, presented in Equation \ref{eq:ESP}, now consists of three components: a gate operations term, a crosstalk term, and a decoherence term.

\begin{equation}
ESP = \prod_{k}\prod_{i}{F_{i,k}}
\label{eq:ESPold}
\end{equation}

\textcolor{black}{where $k$ represents the $k$th time step and $i$ the $i$th gate in the $k$th time step.  $F$ is the fidelity of the corresponding gates.}

\begin{equation}
ESP = (\prod_{k}\prod_{i}{F_{i,k}})\cdot( \prod_{k}\prod_{i,j}C_{i,j,k} )\cdot \textcolor{black}{ (e^{(-t/T_{2}^*)})^N}
\label{eq:ESP}
\end{equation}

\textcolor{black}{where $C$ denotes the crosstalk between the $i$th and $j$th gates during the $k$th time step.} The last term, adapted from \cite{helsen2018quantum}, introduces the decoherence-induced errors, which represent the probability of all qubits staying coherent during the execution of the circuit's execution. $T_{2}^*$ is the decoherence time, $t$ the total duration time of the circuit, and $N$ the total number of qubits. We will discuss each of the terms in detail in the next three subsections.


\subsection{Gate operations term} \label{Operations term}

The selection of typical operational fidelities for each gate, denoted by the \( F_{i,k} \) term, plays a significant role in determining the success probability of quantum circuits. In our analysis, we consider these fidelities to be constants within the design space, rather than variables. This is because determining the optimal fidelity value for increasing the figure of merit (i.e., ESP) is straightforward; higher operational fidelity invariably leads to improved performance. Since we are conducting an exploration, we take a highly optimistic approach to the state-of-the-art \cite{philips2022universal,xue2022quantum,noiri2022fast} values for each gate type\footnote{Another reason we do so is to avoid \textit{numeric underflows} where there is a loss of accuracy in numerical calculations if ESP becomes smaller than the smallest positive representable value in a programming language's floating-point arithmetic.}:

\begin{equation}
    F_{\text{single-qubit}} = 0.9999, F_{\text{two-qubit}} = 0.9998
\end{equation}


\subsection{Crosstalk term} \label{Crosstalk term}

Crosstalk in spin qubit systems has received limited attention in the literature, primarily due to the complexity of accurately modeling the effects that arise when multiple gates are executed simultaneously in close physical proximity \cite{undseth2023nonlinear,heinz2021crosstalk,jirovec2025exchange}. Nonetheless, the presence of crosstalk is expected to have a large influence on quantum computer performance in the future, and current proposals are already suggesting techniques to mitigate it in simple occurrences \cite{undseth2023nonlinear}. For our purposes, crosstalk creates necessary trade-offs between architectural variables, and, as explained before, this is essential for a fair ESP calculation across architectures. For these reasons, we have devised a model to help us discover relative categorical differences between architectures, rather than accurate predictions of crosstalk effects. Our overall aim is not to approximate the actual performance of designs but to pronounce their differences by creating a performance order between them. We further assume that crosstalk should depend on local hardware characteristics, and since this information is already included in the operational fidelity of each gate type, our crosstalk definition is a function of those. Below, we summarize our assumptions:


 \begin{itemize}
     \item Nearest-neighbor crosstalk only (i.e., a pair of qubits that are directly connected by an edge in the coupling graph)
     \item Crosstalk occurs for each edge connecting qubits, which are operated in parallel by different gates
     \item There should be some correlation between the fidelity of gates and crosstalk effects, as they are both influenced by the same quality of fabrication
     \item The crosstalk occurring between neighboring qubits on which two different quantum gates are applied should not be higher than the combined operational error of these two gates
 \end{itemize}

Based on these assumptions, we calculate the crosstalk effect as follows:

\begin{equation}
    C_{i,j,k} = 
       {(\frac{2}{\frac{1}{F_{i,k}}+\frac{1}{F_{j,k}}})}^n
    \label{eq:cr2}
\end{equation}

where $n$ is the number of the total direct links (i.e., nearest-neighbors) on the topology between the $i$th and the $j$th operation. Similarly to before, $F$ corresponds to the operational fidelity.

\begin{figure}[htpb]
    \centering
    \includegraphics[width=1\linewidth]{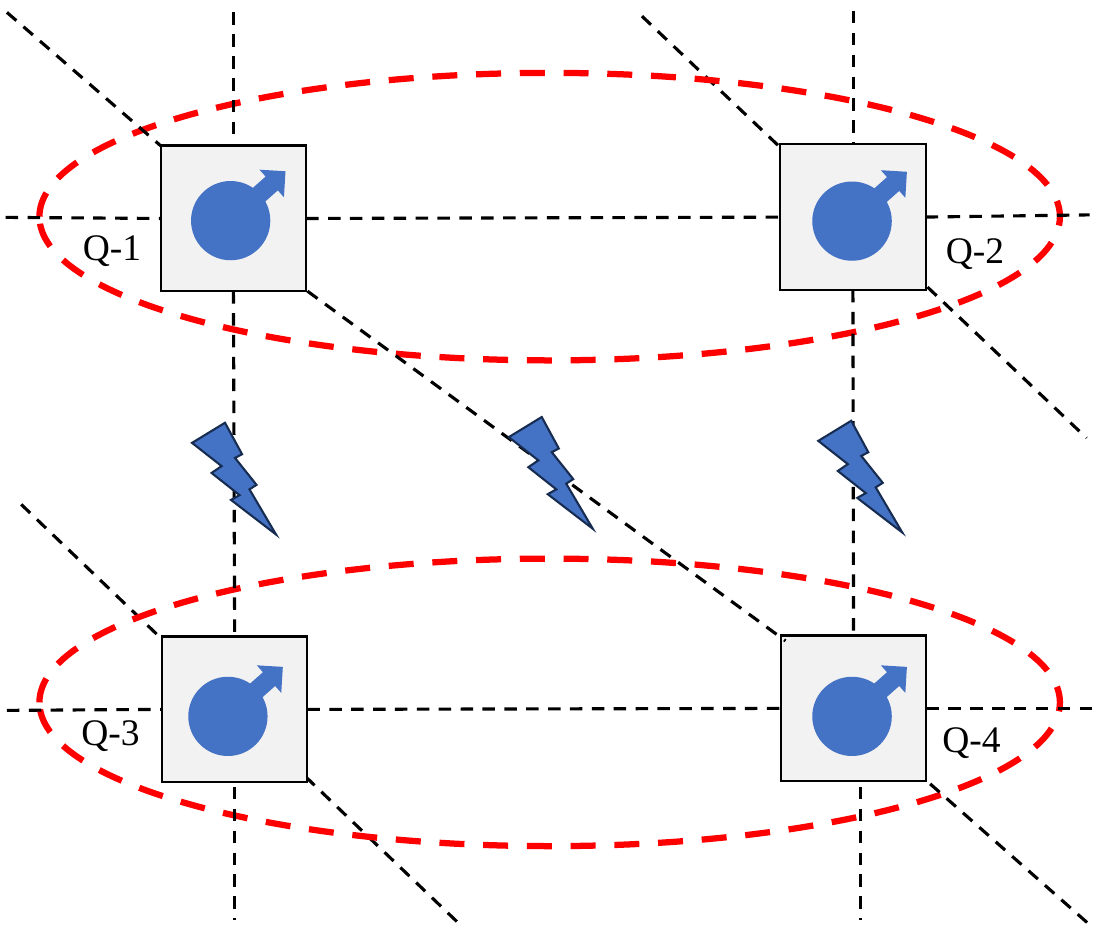}
    \caption{Schematic overview of crosstalk during the execution of two two-qubit gates. Dashed black lines represent direct qubit connections, red ovals indicate active two-qubit gates, and blue lightning icons denote edges where crosstalk effects occur.}
    \label{fig:cr}
\end{figure}

A schematic overview of crosstalk effects is presented in Figure~\ref{fig:cr}. The diagram illustrates two two-qubit gates, represented by red ovals, acting between qubits Q-1 and Q-2, and Q-3 and Q-4, respectively. In this hypothetical architecture with \texttt{degree} = 4 connectivity, crosstalk arises along the three edges marked with a lightning symbol, as the involved qubits are directly connected to one another.

\subsection{Decoherence term}

The last term of Equation \ref{eq:ESP} represents the decoherence of quantum information, which creates a strong architectural trade-off between parallelization and crosstalk effects. Since $N$ and $T_{2}^*$ are given, we need to define the duration time of the circuit $t$:

\begin{equation}
    t = \sum_{k=1}^M\text{max}(D_{i,k})
    \label{eq:ex}
\end{equation}

where $D_{i,k}$ is the duration of the $i$th operation in the $k$th time step. In Table \ref{tab:durations}, we present the gate duration and decoherence time used. The motivation behind this selection of durations, once again, is for purposes of architectural comparisons. We took reference for our two-qubit gate (i.e., $\sqrt{SWAP}$) from \cite{noiri2022fast,xue2022quantum}, for the $shuttle$ from \cite{helsen2018quantum} which is half the shuttle-based \textit{Z} rotations \cite{SpinQ}, for the pulse-based single-qubit gates from \cite{philips2022universal}, and for \textit{$T_{2}^*$} from \cite{hendrickx2021qubit}.

\begin{table}
\centering
    \caption{Overview of all gate duration and the decoherence time used.}
    \includegraphics[width=1\linewidth]{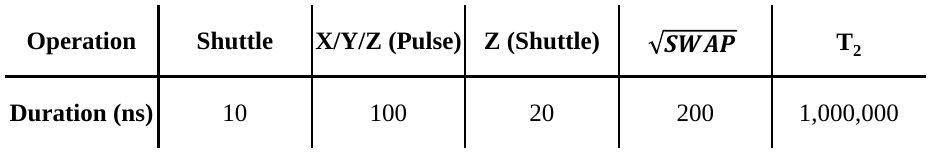}
    \label{tab:durations}
\end{table}

\section{\textit{ArtA} -- \textbf{Art}ificial \textbf{A}rchitect} \label{ArtA (Artificial Architect)}

From the aforementioned transformation of the \textit{SpinQ} compilation framework to a fully functional DSE framework, we are able to brute-force through all valid combinations of the input variable values. However, as we discussed, this can be a time-consuming process, especially when the defined design space is large. Therefore, a specialized tool is needed to automate the DSE process by traversing the space in a clever and faster way. As a result, fewer design iterations will be tried to achieve the same ESP (range) or other specified performance metrics.

\begin{figure*}
    \centering
    \includegraphics[width=1\textwidth]{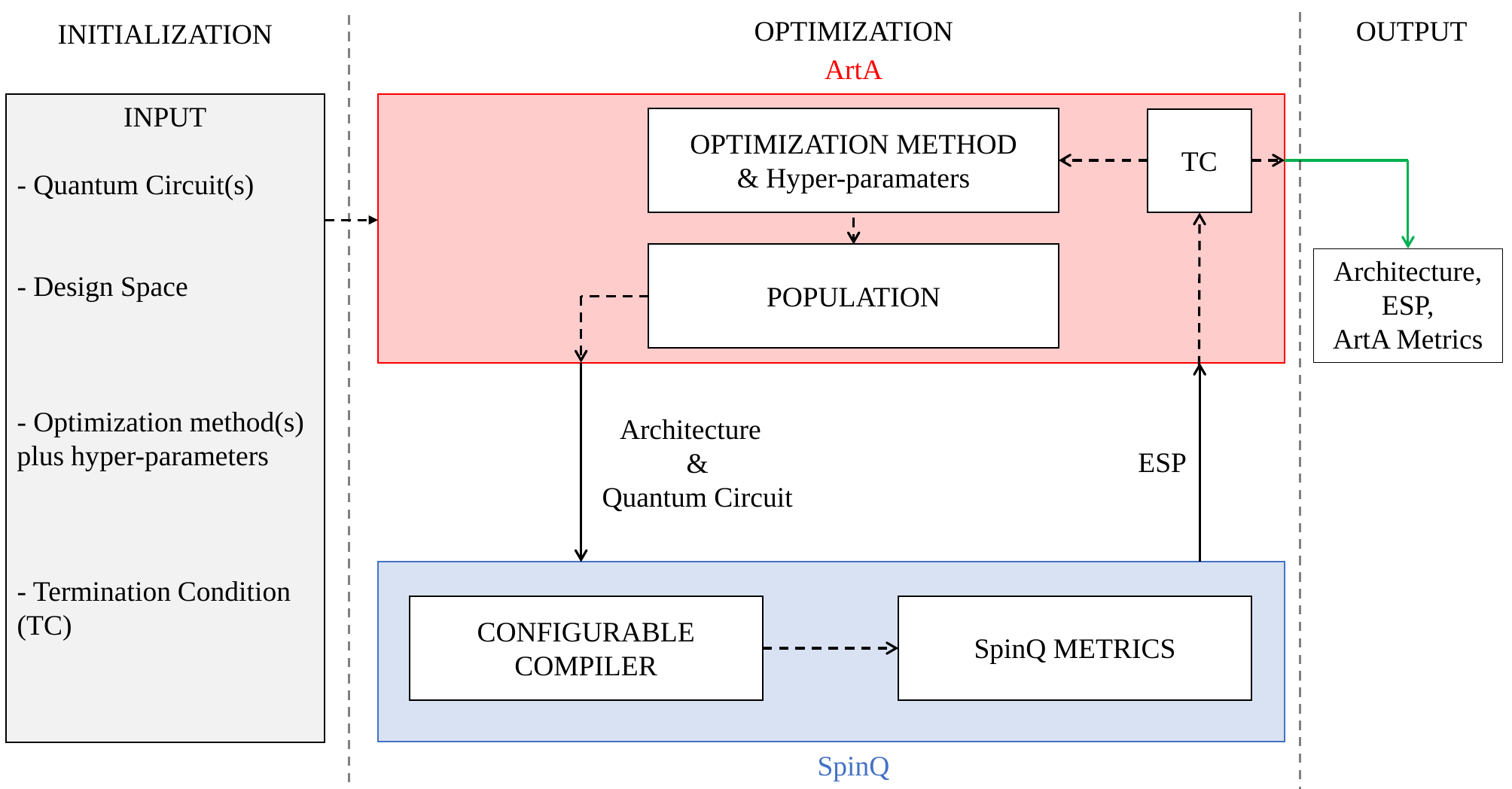}
    \caption{Schematic overview of the initialization, optimization, and output cycle when using ArtA (red box) together with the upgraded \textit{SpinQ} framework (blue box).}
    \label{fig:arta}
\end{figure*}

To do that, we introduce \textbf{ArtA}, a tool integrated into the \textit{SpinQ} framework that incorporates multiple optimization techniques. Its purpose is twofold: (a) to determine which optimization method is most effective at identifying architectures that achieve the desired ESP for each quantum circuit, and (b) to identify the architecture that yields the highest ESP both on a per-circuit basis and across all circuits collectively (i.e., a universal architecture).



The abstracted ArtA process is depicted in Figure \ref{fig:arta}. The initialization of ArtA starts with one or more quantum circuits, a design space such as the one described in Section \ref{SpinQ and the design space}, the optimization method(s) to traverse the space, and an (optional) termination condition. Inside the optimization stage, ArtA provides an architecture and circuit to \textit{SpinQ} for compilation, gets the ESP value, and, based on a policy of the optimization method, selects the next architecture in the form of input variable values from the available ``population" of architectures -- a subset of design space selected by the optimization algorithm itself. This process repeats until a specific termination condition (TC) (e.g., elapsed time) is met or a desired range/value of a specific metric is reached (ESP in this work). Upon termination, the resulting architecture, its ESP calculation, and ArtA metrics are stored. In the following sections, we will describe the TCs and ArtA metrics in more detail. 

In the Appendix \ref{Optimization Methods}, we analytically discuss the five optimization methods used and their hyper-parameter variations, totaling seventeen different optimization configurations. Also, in the Appendix \ref{Quantum Circuits}, we present the seven benchmark quantum circuits and their different sizes in terms of qubit counts.

To answer the last two research questions, we will use ArtA's (a) and (b) capabilities. With the first one, we will answer the third research question, in Section \ref{Comparison of optimization methods}, by testing all the implemented optimization methods to determine which finds the highest ESP the fastest for each quantum circuit. Therefore, first we need to compile each circuit for all $29,312$ architectures and store their ESP for later use. Then, we can run each method and determine the methods' performance based on two new metrics introduced in Section \ref{Metrics Evaluating ArtA}, evaluating ArtA's overall performance. With the second capability of ArtA, we will only use the best-performing optimization method, but expand our circuit selection to concretely answer the last research question, in Section \ref{Architecture Analysis}.

\subsection{Termination condition}

During each optimization cycle, a specific TC is checked to reduce further computational time based on which way, (a) or (b), ArtA is used. Since we first obtain the ESP values for all architectures during the third question, as mentioned before, the highest ESP is known and can work as a TC. Additionally, we observed in practice that most runs reach the highest ESP sooner than 40 minutes; hence, this can be another TC. More specifically, our analysis revealed that over 95\% of processes are completed earlier, while a minority extends significantly longer. Given that our evaluation focuses on runtime, omitting these prolonged instances won't affect the overall assessment, as their inclusion would lead to unfavorable evaluations regardless.

Therefore, the TC when answering the third research question is:
\begin{itemize}
    \item If the highest ESP value is found.
    \item If the method's run time exceeds 40 minutes.
\end{itemize}

For the second use of ArtA, which answers the last research question, we investigate the returned architectural designs by using the best optimization method derived from answering the third question. As we will see later in Section \ref{Comparison of optimization methods}, we find that the corresponding chosen optimization algorithm did not exceed 23.4\% of the total $29,312$ architectures in the worst case before finding the one with the highest ESP. 

Therefore, when answering the last research question, the TC triggers:
\begin{itemize}
    \item When the number of evaluated architectures exceeds 23.4\% of the total $29,312$. 
\end{itemize}


\subsection{Metrics evaluating ArtA} \label{Metrics Evaluating ArtA}

To determine the performance of ArtA, in the context of replying to the third research question, we need to define some relevant metrics. The optimization methods that will eventually determine ArtA's performance should return a solution faster compared to searching with the native brute-force approach of \textit{SpinQ}. Otherwise, ArtA will be of no practical interest. We thus use the relative \textit{time-to-solution} as a metric in Equation \ref{eq:rtt}. The numerator includes the total runtime of both the optimization method and compilation until the desired result is reached (or the TC is triggered), and the denominator refers to the total runtime to brute-force the entire design space.

\begin{equation}
    \text{time-to-solution}_{\text{relative}} = \frac{\text{time-to-solution}}{\text{T}_{\text{Brute-force}}} \cdot 100
    \label{eq:rtt}
\end{equation}

\textcolor{black}{When the average compilation time increases due to larger quantum circuits, the execution time of the optimization method becomes relatively less significant in the total runtime. This happens because, as circuits grow in size, the compilation time per architecture dominates the total runtime, while the time taken by the optimization method remains approximately constant regardless of circuit size. Therefore, \textit{time-to-solution} is not sufficient in evaluating ArtA's performance, as it becomes increasingly influenced by compilation time, rather than the efficiency of the optimization method itself. To address this, we also introduce a relative \textit{calls-to-solution} metric in Equation \ref{eq:rcc}.} With \textit{calls-to-solution} in the numerator, we refer to the total number of times \textit{SpinQ} is called to compile for each new architecture. In case the optimization method attempts the same architecture twice or more, the circuit compilation will be skipped since the results are already stored from the first call. Finally, the denominator represents the brute-force approach and is the total number of architectures in the design space, which essentially equals $29,312$.

\begin{equation}
    \text{calls-to-solution}_{\text{relative}} = \frac{\text{calls-to-solution}}{N_{\text{architectures}}= 29,312} \cdot 100
    \label{eq:rcc}
\end{equation}

\section{Results and evaluation} \label{Results and Evaluation}

In this section, we will attempt to completely answer the third and fourth research questions. In both questions, we will focus on finding the architecture that obtains the highest ESP among all within the design space. Since all optimization methods are inherently probabilistic, multiple runs are required to create robust conclusions. Due to limited computational resources\footnote{All time metrics were retrieved by running ArtA with Python 3.10.8 single-threaded on a 2017 MacBook Pro, using a 3.1 GHz Quad Core Intel i7 processor and 16 GB of RAM.}, we ran each method for each hyper-parameter combination ten times alongside the TCs. Besides the \textbf{mean} performance, we will also report the \textit{worst} performing results for reference.  

\subsection{Comparison of optimization methods} \label{Comparison of optimization methods}

\begin{table*}
    \centering
        \caption{Relative \textit{time-to-solution} metric (the lower, the better) for finding the architecture with the highest ESP from the design space. A minus symbolizes that there was at least one run (out of the ten) in which the top ESP value was not reached in time (TC = 40 minutes). These results are colored red, as well as runs where the time taken exceeds the brute-force time (values over 100). The green cells indicate the top five configurations of optimization methods for a given quantum circuit. The \textbf{mean} and \textit{worst} performing configurations from a sample of ten runs are shown for each quantum circuit (block of rows). \textbf{QFT}: Quantum Fourier Transformation, \textbf{cA}: Cuccaro Adder, \textbf{vA}: vbe Adder, \textbf{QV}: Quantum Volume, \textbf{G}: Grover's, \textbf{RC}: Random Circuit, \textbf{RS}: Random Sampling, \textbf{SA}: Simulated Annealing (Top: Start Temperature, bottom: Step Size), \textbf{BO}: Bayesian Optimization (Top: Acquisition function, bottom: Kernel), \textbf{GA}: Genetic Algorithm (Top: Population Size, bottom: Mutation), \textbf{ACO}: Ant Colony Optimization (Top: Population Size, bottom: Exploitation Rate).}
    \includegraphics[width=1.0\linewidth]{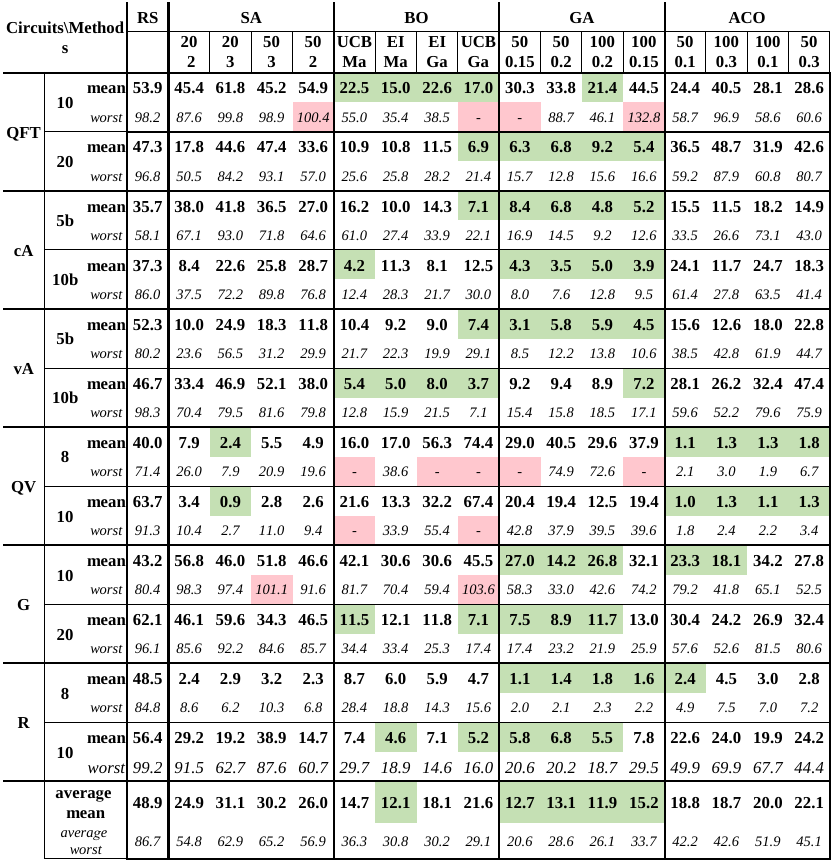}
    \label{tab:time0}
\end{table*}

In this first part of our results, we will address the third research question stated in Section \ref{Problem Statement}, which is about finding the best performant optimization method according to our ArtA metrics defined in Section \ref {Metrics Evaluating ArtA}. The results of the relative \textit{time-to-solution} metric for finding the architecture with the highest ESP are given in Table \ref{tab:time0}. This table also provides insights into matching different optimization method configurations (described in Appendix \ref{Optimization Methods}) with specific quantum circuits. We observe overall that for each quantum circuit, on average, all optimization methods find the solution in less time than brute-forcing (values under 100). For a better interpretation of the numbers on the table, a 50 means the particular optimization configuration of ArtA takes half the time of brute-forcing all architecture, and 100 means that it takes the same time.


Multiple optimization methods greatly outperform the random sampling (RS column), as expected. The best performance is at \textbf{0.9}\% on average for QV10 with SA [20, 3], meaning that ArtA, with this optimization method configuration and quantum circuit, takes only \textbf{0.9}\% relative \textit{time-to-solution}, on average, of the total brute-force time to find the architecture with the highest ESP.

We calculated based on Table \ref{tab:time0} that, on average, for all optimization methods, ArtA was able to decrease the exploration time by \textbf{80.6}\%. Overall, different optimization methods clearly show varied performance across their respective columns. However, the performance differences among various hyper-parameter settings within a single optimization method are less distinct. In a performance order, we can observe GA being first, followed by BO, ACO, and, lastly, SA. More specifically, GA, with a population size of 100 and a mutation probability of 0.2, finds the highest ESP the fastest at \textbf{11.9}\% relative \textit{time-to-solution}, on average, with BO [EI, Ma] coming close at \textbf{12.1}\%. Lastly, in \textbf{86.7}\% of the cases, a top 5 optimization method configuration in the \textit{worst} row is also a top 5 optimization method configuration in the \textbf{mean} row, which indicates the consistency of the results across the ten runs. To further support our findings, next, we will assess the relative \textit{calls-to-solution} metric.





The results of the relative \textit{calls-to-solution} metric for reaching the highest ESP are given in Table \ref{tab:call0}. We observe overall that for each quantum circuit, all optimization methods were able to find the highest ESP architecture in fewer compiler calls than brute-forcing through all $29,312$ architectures. Once again, for a better interpretation of the numbers on the table, 50 means the particular optimization configuration of ArtA needs to go through half the architectures in the design space, and 100 means it needs to go through all the architectures. 

The best optimization method only takes \textbf{1}\% of the total compiler calls, on average, for QV8 and QV10 with ACO [50, 0.1]. We calculated based on Table \ref{tab:call0} that, on average, ArtA's optimization method configurations can decrease the compiler calls by \textbf{79.9}\%. We can observe, once more, that BO, followed by GA, obtained the best performance compared to other optimization methods. Lastly, in \textbf{91.7}\% of the cases, a top 5 optimization method configuration in the \textit{worst} row is also a top 5 optimization method configuration in the \textbf{mean} row, showing results consistency again.

\begin{table*}
    \centering
    \caption{Relative \textit{calls-to-solution} metric (the lower, the better) for finding the architecture with the highest ESP from the design space. A minus with red color cells symbolizes that there was at least one run (out of the ten) in which the top ESP value was not reached in time (TC = 40 minutes). The green cells indicate the top five configurations of optimization methods for a given quantum circuit. The \textbf{mean} and \textit{worst} performing methods from a sample of ten runs are shown for each quantum circuit (block of rows). \textbf{QFT}: Quantum Fourier Transformation, \textbf{cA}: Cuccaro Adder, \textbf{vA}: vbe Adder, \textbf{QV}: Quantum Volume, \textbf{G}: Grover's, \textbf{RC}: Random circuit, \textbf{RS}: Random Sampling, \textbf{SA}: Simulated Annealing (Top: Start Temperature, bottom: Step Size), \textbf{BO}: Bayesian Optimization (Top: Acquisition function, bottom: Kernel), \textbf{GA}: Genetic Algorithm (Top: Population Size, bottom: Mutation), \textbf{ACO}: Ant Colony Optimization (Top: Population Size, bottom: Exploitation Rate).}
    \includegraphics[width=1.0\linewidth]{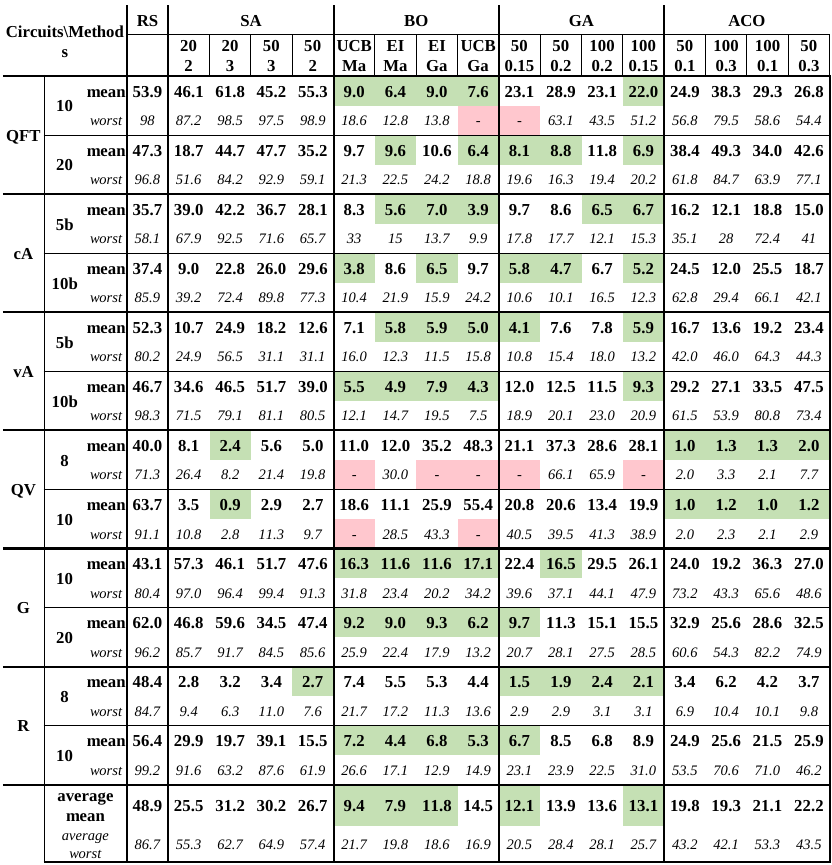}
    \label{tab:call0}
\end{table*}

From the different variants of the GA and BO used, we focus on GA [100, 0.2] and BO [EI, Ma], as they have the best performance and did not reach our TC in any of the ten runs. By comparing the two tables, we can observe that BO [EI, Ma] needs marginally more time to evaluate fewer architectures compared to GA [100, 0.2]. Since the compilation time per architecture is the same between methods, this points to a slightly (0.2\% on average) longer execution time of BO [EI, Ma] compared to GA [100, 0.2]. This can be explained based on the algorithm's reliance on matrix multiplication and several other linear algebra operations, which are computationally more expensive than the more simple operations of GA (see Appendix \ref{Optimization Methods}). However, the actual total runtime of ArtA will depend on the ratio between compilation and optimization time, which can change for large circuits. Based on their small relative \textit{time-to-solution} difference, we can predict that BO [EI, Ma] will outperform GA [100, 0.2] for larger circuits as the latter has, on average, 58\% higher relative \textit{calls-to-solution}. Therefore, we conclude that BO [EI, Ma] is a better candidate for automating the DSE process for large-scale circuits, as it can traverse the design space faster. 

Having said that, making a more detailed analysis for each circuit where internal circuit characteristics are correlated \cite{SpinQ,bandic2023interaction} to optimization methods can improve the performance of ArtA by picking methods more accurately. This promising avenue is left for future work when more progress in circuit characterization is made.

\subsection{Architecture analysis} \label{Architecture Analysis}

\begin{table*}
    \centering
        \caption{Architectures returned by BO [EI, Ma] optimization method for a range of quantum circuits (refer to Appendix \ref{Quantum Circuits}). Results are color-coded, with color schemes following the categories introduced in Section \ref{SpinQ and the design space}. Final-row averages omit -1 values.}
    \includegraphics[width=1.0\linewidth]{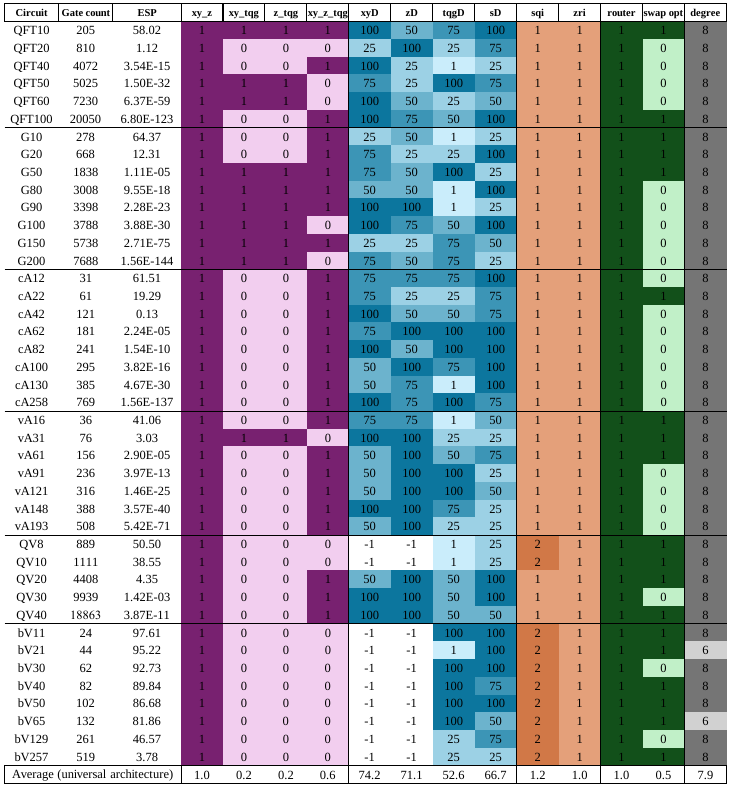}
    \label{tab:archs}
\end{table*}


We move on to address the last research question after having selected the best-performing optimization method in the previous section. Here, we use ArtA to create architectural insights into spin-qubit devices for each circuit category and suggest a universal architecture for all executed circuits.

In Table \ref{tab:archs}, we present the values of the architectural variables, introduced in Section \ref{SpinQ and the design space}, that were found by ArtA having the highest ESP with the BO [EI, Ma] optimization method alongside their respected ESP values. These circuits are organized by class and number of qubits within the rows of the table. Subsequent columns present the ESP of the best architecture discovered, followed by color-coded architectural parameter categories as delineated in Section \ref{SpinQ and the design space}. The use of color accents in the table serves to enhance visual clarity, with each corresponding to the values of each variable. 

Before focusing on the architectural insights, we should comment on the low ESP values observed, especially in larger circuit sizes. Although such low values do not have any physical meaning after a certain small number, ESP still remains a reliable way to rank architectures. It's important to understand ESP is not inherently random, regardless of how low it gets. For instance, slight variations between two architectures causing a single gate difference will consistently and reliably be reflected through their ESPs. As long as numeric underflows are not caused, ESP remains a robust figure of merit for our purposes. 

Observing the data presented in Table \ref{tab:archs}, it becomes evident that there is a discernible structure to the values of architectural variables. Notably, many or all circuits exhibit strong preferences for certain variables. Following an implications discussion on the circuit performance of these strong architectural preferences, we will explore areas of weaker preference where there is more variability in value choices.


\textbf{Strong preference} 

First, we observe a unanimous preference for the Z rotation implementation (\texttt{z\_rot\_impl}) in the \textbf{zri} column, with a pulse-based rotation being preferred. Note that previously, we have assumed in Section \ref{Operations term} that a shuttle and a pulse-based Z rotation share the same single-qubit gate fidelity, but because the latter requires two shuttles, it can benefit the ESP more than the other type. Moreover, as a shuttle operation involves two quantum dots while a pulse-based operation only involves one, the shuttle-based Z rotation will have a higher chance of incurring crosstalk errors based on our crosstalk definition. Given these two reasons, BO has favored the pulse-based version. This indicates that shuttle-based single-qubit rotations are not sustainable compared to pulse-based ones for large-scale circuits as they progressively contribute more noise.

Moving on, since the X, Y, and Z gates are implemented the same way hardware-wise, \texttt{xy\_z} automatically is assumed 1 based on the interdependencies detailed in Section \ref{Design interdependencies and verification}. Consequently, \texttt{xy\_tqg} and \texttt{z\_tqg} are equivalent, taking both either 0 or 1 from the optimization method of ArtA (i.e., BO [EI, Ma] in this case), as we can observe in the table. Additionally, \texttt{xy\_z\_tqg} can also be conceptually equivalent to \texttt{xy\_tqg} and \texttt{z\_tqg} since pulse-based Z rotations are selected. Therefore, whenever \texttt{xy\_z\_tqg} is 1 or \texttt{xy\_tqg} and \texttt{z\_tqg} is 1, it means the architecture(s) which allow(s) parallelization of all single- and two-qubit gates obtains the highest ESP. In that sense, one would expect \texttt{xy\_z\_tqg}, \texttt{xy\_tqg} and \texttt{z\_tqg} to be synchronized. However, since such exploration freedom was given in the design space, the reasons for the mismatches could be explained by the internal characteristics of the circuits themselves. For example, in QFT100, the parallelization of all gate types occurs more often than in QFT60, as observed from their gate parallelization constraints \texttt{xy\_z\_tqg} = 0, \texttt{xy\_tqg} = 1, and \texttt{z\_tqg} = 1. 

Conversely, when applying the same reasoning to QV8, QV10, and QFT20, it appears that hardware support for parallel execution of single- and two-qubit gates is not essential. This highlights that the success of real quantum circuits can stem from different architectural “angles,” even under reduced hardware capabilities \cite{near-term}. This exemplifies the strength of our approach: it enables the identification of optimal architecture configurations without requiring the most complex or fabrication-intensive hardware designs. \footnote{Ultimately, each circuit profile has specific architectural traits that can be leveraged to achieve optimal performance without the need for excessively advanced hardware features (achieving a balance in architectural characteristics).}

Another strong preference appears in the \textit{beSnake} router, observed by the dark green 1 values in the \textbf{router} column. This result shows that the crosstalk implications of using the \textit{beSnake} router (which gives more shuttling freedom) over the \textit{shuttle-based SWAP} router (which minimizes crosstalk) are not as important as the other two components in Equation \ref{eq:ESP}. Similarly, the \texttt{degree} variable is mostly maximized, indicating again that the additional possibility of crosstalk does not outweigh other benefits, such as having more connections to perform two-qubit gates and shuttling operations. 

Lastly, there is a strong preference for a local single-qubit gate implementation (\texttt{single\_qubit\_impl}). A local implementation, on the one hand, provides the highest parallelization possibilities, but on the other, it is the most demanding to achieve experimentally on real hardware. 

Based on all the above conclusions, we can summarize that the crosstalk effect is not as strong even though ArtA suggests architectures with high crosstalk factors (the highest connectivity and parallelization and the most dense routing strategy). In fact, the decoherence-induced errors increase more severely (exponentially) with time, based on Equation \ref{eq:cr2}, than crosstalk, especially in circuits with high qubit and gate counts, which explains the prevailing preference for maximizing parallelization.

The same can be observed in our universal architecture recommendation in the last row when rounding the numbers, which performs best on average for all the circuits examined. Additionally, the operational fidelity, which is directly linked to the gate overhead of compilation, can also severely impact the final ESP for circuits with high gate counts, which in turn explains the need to add the least additional gates possible during compilation through the use of \textit{beSnake} \cite{paraskevopoulos2024besnake} and local single-qubit implementation. Having said that, Bernstein Vazirani, QV8, and QV10 circuits preferred the global implementation, as these circuits consist of a significant number of single-qubit gates (with the same angles) being executed at the same time step.

\textbf{Weak preference}

Overall, we can observe that the single-qubit operations and shuttling prefer a higher degree of parallelization than the two-qubit gates. Considering that \texttt{xyD} and \texttt{zD} have the same meaning due to \textbf{zri} = 1, their combined averages from the last row of Table \ref{tab:archs} is \textbf{72.7}\%. This compares to an average of \textbf{52.6}\% for the two-qubit gates and \textbf{66.7}\% for \texttt{sD}. Therefore, prioritizing the simultaneous operation of single-qubit gates (and shuttling) over two-qubit gates is more important.

The \texttt{swap\_opt} variable seems to show a preference for supporting SWAP gates for lower qubit counts. A possible explanation for this is that SWAP operations take a longer time than a parallelized sequence of shuttles \cite{paraskevopoulos2024besnake}. As we explained before, the effects of having longer circuit execution times become more severe with larger circuits; hence, the extra SWAPs can outweigh the gained fidelity they initially offer. Combining this conclusion for large-scale circuits with the observed strong preference for shuttling flexibility, enabled by the beSnake router, suggests that shuttle operations constitute a key communication mechanism in spin-qubit architectures for the future.


\section{Conclusion} \label{Conclusion}

In this paper, we propose a comprehensive exploration of the current and future architectural design space for quantum dot spin-qubit quantum processors. Our research focuses on assessing the critical architectural characteristics that could be key in ensuring high performance on such devices. To this end, we present an upgraded version of the \textit{SpinQ} compilation framework \cite{SpinQ} for spin qubit architectures in which architectural variables can configure internal compilation passes, resulting in the first Design Space Exploration (DSE) framework. After defining our design space, consisting of $29,312$ architectures, we propose a multi-optimization-based tool, ArtA (\textbf{Art}ificial \textbf{A}rchitect), to automate the DSE process. The goal of \textbf{ArtA} is twofold: (a) to identify which of the seventeen optimization configurations can most efficiently discover the architecture that achieves the highest ESP, compared to brute-force exploration; and (b) to determine the architectural design characteristics that are critical for constructing high-performance spin-qubit devices across different circuit categories. We have shown up to \textbf{99.1\%} improvement in computation times compared to brute-forcing, showing that it is possible to easily explore a vast number of designs. We have also provided insights into matching different optimization methods with specific quantum circuit categories. 

\textcolor{black}{It should be noted, that the underlying goal of our study was not to exhaustively benchmark all possible optimization methods, but rather to introduce the concept of applying design space exploration to quantum computing systems -- with a particular focus on spin-qubit architectures -- by employing a diverse and representative selection of methods from different optimization families.} Bayesian optimization with an Expected Improvement acquisition function and a Matérn kernel was chosen as the best method due to its effectiveness in managing higher qubit counts. After that, we proceed to automate the DSE process for 44 quantum circuits of up to 258 qubits within the 29,312-architecture design space. Our findings reveal the importance of minimizing the circuit duration by maximizing parallelization whenever possible instead of selecting architectural characteristics that minimize crosstalk errors. Crosstalk is currently a significant concern in experiments, but our long-term considerations deem decoherence more crucial. Lastly, the shuttle operation is preferred over costly SWAP operations for communication purposes, especially for large-scale circuits. Contrary to that, shuttle-based single-qubit gates such as Z rotations are not preferred over pulse-based. Furthermore, our universal architecture suggestion places higher priority on parallelizing single-qubit gates rather than two-qubit gates. Overall, our work demonstrates how the integration of DSE methodologies with optimization algorithms can uncover valuable architectural insights that would be difficult, if not impossible, to foresee while suggesting points of optimality with reduced hardware complexity.

\textcolor{black}{However, it is important to acknowledge that while these findings highlight promising architectural directions, several of these design choices -- such as high degrees of parallelization, connectivity, and flexible shuttling strategies -- currently pose significant engineering challenges. Implementing fully local control, for instance, demands high fabrication uniformity and precise calibration, which is non-trivial at larger scales. Likewise, extensive shuttling flexibility requires precise individual gate addressing, which is experimentally demanding. Nonetheless, the purpose of this work is to chart a clear path forward, identifying architectural features that, while ambitious, can guide future efforts toward scalable and performant spin-qubit processors.} 

As for future improvements, a broader range of hyper-parameter combinations and circuit variations could enhance ArtA's effectiveness. Then, with advancements in spin qubit device fabrication, new architectural possibilities could potentially expand the design space, and such integration will be relatively straightforward owing to the modularity of the internal functions of \textit{SpinQ} and ArtA. Moreover, introducing non-architectural constraints such as development costs and time into ArtA's design space could help make real-world predictions regarding developing spin qubits. For instance, this would allow for operational fidelities to become variables.

Also, some of the assumptions made in this work could be fine-tuned as the field progresses. It is crucial to acknowledge the impact of these assumptions on the observations presented. For instance, we assumed the fidelity of pulse-based Z rotations to be equivalent to that of a single shuttle operation. This assumption should not be construed as either favorable or unfavorable; rather, it is a foundational premise upon which ArtA bases its architectural ``recommendations". Should empirical evidence later demonstrate that shuttle operations exhibit significantly higher fidelity compared to pulse-based Z rotations or the crosstalk model depends on more factors, it may necessitate a reevaluation of preferences.

Further analysis of architectural patterns associated with specific quantum circuit characteristics may reveal additional trends and enable systematic categorization. For example, circuits with a high proportion of two-qubit gates (e.g., 90\% or more) may exhibit optimal performance only when the qubit connectivity exceeds a certain threshold. Insights like these, based on various circuit properties \cite{bandic2023interaction,SpinQ}, can inform the construction of a generalized lookup table that maps algorithmic profiles to architectural configurations. Such a resource would enable rapid prediction of high-performing architectures for new circuits, eliminating the need to re-run the full DSE process.

\section{Acknownledgement}
This work is part of the research program OTP with project number 16278, which is (partly) financed by the Netherlands Organisation for Scientific Research (NWO). CGA acknowledges funding from the Spanish Ministry of Science, Innovation and Universities through the Beatriz Galindo program 2020 (BG20-00023) and from the European ERDF under grant PID2021-123627OB-C51.

\section*{Declarations}

There are no competing interests between the authors. 

\newpage

\bibliographystyle{ACM-Reference-Format}
\bibliography{bibliography}

\section{Appendix} \label{Appendix}

\subsection{Quantum circuits} \label{Quantum Circuits}

\begin{table*}
    \centering
    \caption{Summary of all quantum circuits analyzed to answer the third and fourth research questions. All qubit counts, followed by a *, are used in answering both research questions, while an absence of a symbol indicates algorithms used in the fourth research question only. A ! indicates circuits \textit{only} used in the third research question.}
    \resizebox{1.0\textwidth}{!}{
    \begin{tabular}{p{3cm}|p{9cm}|p{5cm}}
        \textbf{Circuit name} & \textbf{Description} & \textbf{Qubit count} \\
        QFT & Produces a highly entangled state including a phase factor \cite{Coppersmith2002RCFactoring} & 10*, 20*, 40, 50, 60, 100 \\
        Grover's search & Used to perform database search \cite{Grover1996ASearch} & 10*, 20*, 50, 80, 90, 100, 150, 200\\
        Quantum Volume & Created to study the Quantum Volume metric \cite{cross2019validating} & 8*, 10*, 20, 30, 40 \\ 
        Random circuits & Circuit used for benchmarking quantum computers & 8*!, 10*!
        \\
        Cuccaro Adder & Used for the addition of two numbers \cite{Cuccaro2004ACircuit} & 12*, 22*, 42, 62, 82, 100, 130, 258 \\
        vbe Adder & Also used for addition, uses more ancilla qubits \cite{Vedral1996QuantumOperations} & 16*, 31*, 61, 91, 121, 148, 193\\
        Bernstein Vazirani & Finds an unknown bitstring \cite{Bernstein1993QuantumTheory} & 11, 21, 30, 40, 50, 65, 129, 257 \\
    \end{tabular}}
    \label{tab:alg}
\end{table*}

Quantum circuits are defined as a sequence of gates to be executed on qubits (single-qubit gates) or between qubits (two- or multiple-qubit gates). Such quantum circuits are often given in a hardware-agnostic form. This means there are no considerations for device-specific restrictions, such as connectivity of qubits or the set of executable gates. Therefore, a compiler \cite{SpinQ,sivarajah2020t,Qiskit,khammassi2021openql,salm2021automating,Developers2023-gs,computing2019pyquil,javadiabhari2014scaffcc,chong2017programming,wu2023intel} is necessary to transform a hardware-agnostic circuit into a circuit that is executable on a given device architecture.

We have selected a wide range of representative quantum circuits of different qubit counts from Qlib \cite{lin2014qlib} and qbench \cite{bandic2023interaction} libraries to thoroughly test \textit{SpinQ} and ArtA. Table \ref{tab:alg} summarizes all the quantum circuits used in this work. 


\subsection{Optimization methods} \label{Optimization Methods}

We introduce the five optimization methods, four of which have four different hyper-parameter configurations, totaling seventeen configurations. \textcolor{black}{Our goal was not to exhaustively benchmark all possible optimization methods, but rather to introduce the overall concept of applying optimization to the DSE process for quantum computing systems, with a focus on spin-qubit architectures, using a diverse and representative selection of methods. More specifically, from the family of nature-inspired evolutionary algorithms, we selected the well-established Genetic Algorithms (GA); from the class of physics-inspired metaheuristics, we chose the widely used Simulated Annealing (SA); from surrogate-based optimization methods, we included Bayesian Optimization (BO); and from the Swarm Intelligence class, we incorporated Ant Colony Optimization (ACO). In the following sections, we} will introduce each method, describe the policy it uses to select a new architecture in each iteration loop of ArtA, give additional details, and finally describe the hyper-parameters used. Several of the optimization methods rely on a notion of distance between architectures. Since our design parameters include categorical variables, we begin by outlining how a meaningful distance metric can be defined in this context

\subsubsection{Distance measure}

The notion of distance is hard to apply to values of categorical variables because a categorical variable is defined by having no order between the possible values. For example, the set $\phi = \{ \text{apple, pear, banana}\}$ is categorical, as the ``$<$" or ``$>$" signs hold no value when applied to this set. Consequently, no order means no distance can be introduced.


There is only one variable that is categorical: \texttt{single\_qubit\_ impl}. To solve this, we postulate an ordering of values by the maximum amount of gates that would be theoretically executed during the same time step. The ordering we used in this work is then \textit{sequential} $<$ \textit{semi-global} $<$ \textit{local} $<$ global. This means, for instance, the distance between \textit{sequential} and \textit{semi-global} is 1, but the distance between \textit{sequential} and \textit{local} is 2.

The second challenge is how to combine the distance measured between values of one variable with the distance measured between values of a different variable. To describe this, we define $d$ as the distance between two values of the same variable and $D$ as the distance between architectures such that the relation between $D$ and $d$ is given in Equation \ref{eq:dis}. This implies that the distance between two architectures is the sum of the distances between the corresponding variable values. Finally, we decided to count all distances in units of 1 and have the distance equal to the difference in indices if we were to order the set of allowed values of one variable. This holds for all values, such that $d(\texttt{degree}=4,\texttt{degree}=6)=1$, $d(\texttt{degree}=4,\texttt{degree}=8)=2$, etc. 

\begin{equation}
    D(\text{arch}_1,\text{arch}_2) = \sum_{\text{arch\_variable}}d_{\text{arch\_variable}}(\text{arch}_1,\text{arch}_2)
    \label{eq:dis}
\end{equation}


\subsubsection{Random sampling}

Random sampling is used as our baseline to compare other optimization methods. It traverses the design space by randomly selecting architectures to investigate. The policy comprises assigning one randomly sampled architecture as the population. 


\subsubsection{Simulated annealing}

SA \cite{Kirkpatrick1983OptimizationAnnealing} is inspired by the metallurgical process of heating and then slowly cooling a material, which minimizes defects and thus approximates its lowest energy state. SA is a mix of a greedy strategy for selecting the best-performing architecture while trying to escape local optima to find the global optimum. A more detailed overview of simulated annealing can be found in \cite{Suman2006AOptimization}.
The policy is as follows:

\begin{enumerate}
    \item Start with a random architecture, called architecture$_{i}$, which has ESP$_{i}$, where $i$ is the iteration number. It should be noted SA only works with a population size of one, which is the architecture$_{i}$.

    \item Set a starting temperature ${T_i} = T_{\text{start}}$ (hyper-parameter set to either 50 or 20).

    \item Find a random candidate architecture$_{c, i}$ which has distance $D_i$ to architecture$_{i}$. We assume that $D_i$ is randomized but remains greater than 0 to avoid picking the same architecture. Then, $D_i = \text{max}(\text{Normal}(1,1) \cdot step\_size,1)$, with $step\_size$ used as a hyper-parameter equal to either 2 or 3, and Normal(1,1) indicates a normal distribution with mean and standards deviation equal to 1. In case no architecture$_{c, i}$ with distance $D_i$ to architecture$_{i}$ is found, a new $D_i$ is generated, and this step repeats until an architecture$_{c, i}$ is picked.

    \item Calculate the selection probability: $C_i = \text{exp}({(\text{ESP}_{c,i} - \text{ESP}_i)}/{T_i})$.

    \item Accept the candidate architecture as current architecture if $C_i > u$, where $u$ is a random number between 0 and 1.
        
    \item Update temperature $T_i$ (linear cooling process) as $T_i = (T_{\text{start}}/i$)

    \item Repeat steps 3-6 until the TC is met.
        
\end{enumerate}

The temperature regulates the probability of accepting a candidate architecture with a lower ESP. Specifically, with higher temperatures, there are higher chances of accepting candidate architectures even if they have worse ESP values. However, as temperature decreases, the algorithm only accepts architectures with higher ESP. This mechanism prevents getting stuck in local maxima at the start of the optimization method while gradually becoming greedy at the end by selecting the top-performing architectures only.

\subsubsection{Bayesian optimization}

BO \cite{frazier2018tutorial,shahriari2015taking} uses conditional probabilities based on previously sampled data points to predict the function value and uncertainty of subsequently sampled data points. 
It typically assumes a Gaussian Process (GP) as the statistical model that governs the relation between the function values of points. 
Our policy for BO is as follows:
\begin{itemize}

    \item We assume the ESP function has been observed at $n_o$ architectures at iteration $i$.
    
    \item The last $n_{\text{sample history}}$ (set to 300 in this work) of the sampled architectures is selected. We call them $\text{a}_s$.
    
    \item From the unsampled architectures, $n_{\text{candidate amount}}$ (set to 1000) architectures are selected. We call them $\text{a}_c$.
    
    \item $\text{a}_s$ is used to build a surrogate function for $\text{a}_c$. The surrogate function predicts the mean and standard deviation for each architecture in $\text{a}_c$. This surrogate function is constructed with a \textit{mean function} $\mu_0$ and a covariance matrix by evaluating a \textit{kernel} $\Sigma_o$. 
    
    \item Given the surrogate function, we compute the acquisition function. This acquisition function uses each architecture's mean and standard deviation in $\text{a}_c$ to determine whether an architecture should be evaluated. It balances a high expected value (exploitation) with a high uncertainty (exploration). As explained later, there are many acquisition functions, each with its own approach to making this exploitation/exploration trade-off.
    
    \item From $\text{a}_c$ we choose $n_{\text{size pop}}$ (set to 50) architectures, namely those with the highest value of the acquisition function.

\end{itemize}
For discrete functions in the BO algorithm, all non-integer valued points are first rounded to integers before calculating their expected value \cite{Garrido-Merchan2020DealingProcesses}. 
Additionally, since not all combinations of architecture variables constitute valid architectures, we use a BO formulation for constrained problems~\cite{Ungredda2021BayesianProblems}. Therefore, we only sample from valid architectures, which automatically satisfy the constraints and only contain the integer values for variables that are of interest to our problem.

For each architecture $a \in \text{a}_c$, our GP assumption lets us compute the conditional distribution of ESP using Bayes' rule~\cite{Frazier2018AOptimization}.
\begin{equation}
\text{ESP}(\text{a})|\text{ESP}(\text{a}_s) \sim \text{Normal}(\mu_s(\text{a}),\sigma_s^2(\text{a}))
\end{equation}
where,
\begin{equation}
    \mu_s(\text{a}) = \Sigma_0(\text{a},\text{a}_s)\Sigma_0(\text{a}_s,\text{a}_s)^{-1}(\text{ESP}(\text{a}_s)-\mu_0(\text{a}_s))+\mu_0(\text{a})
\end{equation}
and,
\begin{equation}
    \sigma_s^2(\text{a}) = \Sigma_0(\text{a},\text{a}) - \Sigma_0(\text{a},\text{a}_s)\Sigma_0(\text{a}_s,\text{a}_s)^{-1}\Sigma_0(\text{a}_s,\text{a})
\end{equation}

For the mean function, we use the arithmetic mean as:
\begin{equation}
    \mu_0(x)|\text{a}_s = \frac{1}{s} \sum_{i=1}^s(\text{ESP}_i)
    \label{eq:mean}
\end{equation}

Two different Kernels are used, namely the Gaussian Kernel and the Matérn Kernel \cite{Frazier2018AOptimization} with $\nu=1.5$, both of which are commonly used for BO problems. 
They are defined as follows:
\begin{equation}
    \Sigma_0(\text{a}_a,\text{a}_b) = \begin{cases}
        \text{Gaussian:} \\
        \hspace{1em}\text{exp}(-||\text{a}_a-\text{a}_b||^2)\\
        \text{Matérn:}  \\
        \hspace{1em}(1+\sqrt{3}||\text{a}_a-\text{a}_b||)\cdot\text{exp}(-\sqrt{3}||\text{a}_a-\text{a}_b||)
    \end{cases}
    \label{eq:ker}
\end{equation}
Given this surrogate function, an acquisition function (AF) is calculated to determine which points will be sampled next.
To this end, we use Expected Improvement (EI) and Upper Confidence Bound (UCB) as AF, defined as:

\begin{align}
    \text{AF}(\text{a}) = \begin{cases}
        \text{EI:} \\
        \hspace{1em}(\mu_s(\text{a}) - \text{max(ESP}_s)) \cdot \text{CDF}(Z)  \\
        \hspace{1em} + \sigma_s(\text{a})\cdot \text{PDF}(Z)\\
        \text{UCB:} \\
        \hspace{1em}\mu_s(\text{a}) + \sigma_s(\text{a}) 
    \end{cases}     
\end{align}

With $Z = (\mu_s(\text{a}) - \text{max(ESP}_s))/(\sigma_s(\text{a}))$.
CDF and PDF are the Cumulative/Probability Density Functions, respectively. 
Variable max(ESP$_s$) equals the highest ESP value found for all architectures present in $\text{a}_s$. 
ArtA then picks the new architectures with the highest acquisition function values to sample in the new iteration until the TC is met. 

\subsubsection{Genetic algorithm}

GA \cite{zbigniew1996genetic} is modeled after the biological evolution of genes to arrive at a near-optimal solution for an optimization problem. The GA changes the architectural variables of high-performing architectures to explore and discover even better-performing architectures with the use of three operations: selection, mutation, and cross-over. The GA policy works as follows: 
\begin{itemize}

    \item Every generation's iteration, $i$, considers a population of size $n$ (hyper-parameter set to either 50 or 100). The population at generation $i+1$ consists of elite, cross-over, and mutation children. These are explained next.
    
    \item From these $n$ architectures, $n_{\text{elite}}$ individuals (set to 15\% of $n$) with the highest ESP value are selected and automatically included in our population of the next generation.
    
    \item Additionally, $n_{\text{parent}}$ (set to 50\% of $n$) out of $n$ architectures are selected iteratively by randomly considering $n_{\text{tournament}}$ architectures (set to 5) from the population and selecting the one with the highest ESP value. There are no duplicate parents.
    
    \item Two parents of this group are selected randomly. Cross-over is performed by picking a random cross-over point $c$ with $1\leq c \leq k$, where $k$ is the total number of architectural variables of the design space. A new architecture is formed by combining the architectural variables of the two parents, with $c$ indicating the index separating the two halves of the variable list. The process is repeated if this is not a valid architecture. $n_{\text{cross-over}}$ (set to 30\% of $n$) architectures are created in this way.
    
    \item The remaining new generation is populated with $n_{\text{mut}}$ architectures (set to 55\% of $n$). For this operation, some children (from before) alter each of their architectural variable values with probability $p_{\text{mut}}$ (hyper-parameter set to either 0.15 or 0.2). The process is repeated if this is not a valid architecture until one is found.

    \item The algorithm repeats until the TC is met. 
\end{itemize}
Note that in GA (as in SA), it is possible to have the same architecture in multiple populations. 

\subsubsection{Ant colony optimization}

The fundamental concept behind the ACO \cite{Dorigo2006AntOptimization} is based on how ants find the shortest route from their colony to a food source. Ants lay down pheromones along their path, and the intensity of the pheromone trail guides other ants to the food source. Over time, the pheromone trail evaporates, reducing its strength. However, the shorter the path, the more frequently traveled by ants, and thus, more pheromone is deposited. This positive feedback eventually leads the colony to converge on the shortest path.

To implement ACO in the context of finding the highest ESP architecture, we need to explicitly pre-create the graph (in this case, a tree) on which ants are walking during the run. The four steps involved are as follows. (i) A ``start" node is created. (ii) After a random valid architecture is selected, the next node after ``start" is created with the first architectural variable value. The subsequent node includes the first and second variable values of this valid architecture. The tree continues to be constructed all the way until the last node contains the variable values. (iii) Sample the rest of the valid architectures of the design space and repeat the same process. Use the same node if the generated node with the same variable values already exists; otherwise, create a bifurcation to a new node. (iv) The ants traverse this tree, and the node at which they end will always be a valid architecture. Having constructed this tree, the policy is as follows:

\begin{itemize}

    \item Pheromone deposition is initialized on all edges of the tree.
    
    \item A population of $n$ architectures (hyper-parameter set to either 50 or 100) is considered at every time step $i$.
    
    \item The tree is traversed $n$ times (each time by an ant) from the top until the end for all $n$ architectures. A selection rule is applied to a node with multiple outgoing nodes based on its pheromone concentration. With probability p$_{\text{exploit}}$ (hyper-parameters set to either 0.1 or 0.3), the outgoing edge with the highest pheromone concentration is taken. Otherwise, exploration is performed with selection probability for each edge, which is the ratio of the pheromone concentration for the edge to the total pheromone amount on the entire tree.

    \item Pheromone concentration on each edge of an ant path is updated with Equation \ref{eq:ph} \cite{Shobaki2022Register-Pressure-AwareOptimization}.

    \item Pheromones decay (set to 10\% rate) for all edges. 
        
    \item The new population at $i+1$ consists of the $n$ architectures these ants arrive at.

    \item The algorithm repeats until the TC is met.
    
\end{itemize}

\begin{equation}
    ph_{ij} = \begin{cases}
        \text{if } \Delta $ESP$_i < 0:\text{\textit{ min\_ph} }   \\
        \hspace{1em}\\
        \text{else if } \Delta\text{ESP\_avg} < 0: \text{\textit{ max\_ph} }  \\
        \hspace{1em}\\
       \text{else}: \text{min} ( \text{\textit{min\_ph}} + (\text{\textit{max\_ph}} - \text{\textit{min\_ph}}) \cdot \frac{\Delta\text{ESP}_i}{\Delta\text{ESP\_avg}},\\
        \hspace{3em} \text{\textit{max\_ph}} )   
    \end{cases}
    \label{eq:ph}
\end{equation}

Here, $ph_{ij}$ is the pheromone deposited on the edge between node $i$ and node $j$. \textit{max\_ph} is the maximum pheromone deposition (set to 6) and \textit{min\_ph} the minimum (set to 1). $\Delta\text{ESP}_i = \text{ESP}_{\text{ant}} - \text{max(ESP)}$, with max(ESP) being the maximum ESP observe so far. $\Delta\text{ESP\_avg} = \frac{1}{3}(\Delta\text{ESP}_{i-1} + \Delta\text{ESP}_{i-2} + \Delta\text{ESP}_{i-3})$. This was included to make large gains in the discovery of high ESP architectures relatively easy at an earlier stage, while at a later stage, increases in ESP will be smaller. Weighing the increase in ESP with this rolling average reflects this.


\end{document}